\documentclass[twocolumn,showpacs,preprintnumbers,superscriptaddress,amsmath,amssymb]{revtex4}

\usepackage{graphicx}
\usepackage{dcolumn}
\usepackage{bm}

\begin{document}

\preprint{APS/123-QED}

\title{Resonant ion-pair formation in electron recombination with HF$^+$}

\author{J. B. Roos}
\affiliation{
Department of Physics, Stockholm University, AlbaNova University Center,\\
 S-106 91 Stockholm, Sweden
}
\author{A. E. Orel}
\affiliation{
Department of Applied Science, University of California, Davis\\
Davis, CA 95616, USA
}
\author{ \AA. Larson\footnote{Corresponding author; e-mail: aasal@physto.se}}
\affiliation{
Department of Physics, Stockholm University, AlbaNova University Center,\\
 S-106 91 Stockholm, Sweden
}

\date{\today}

\begin{abstract}
The cross section for resonant ion-pair formation in the collision of low-energy electrons with
HF$^+$ is calculated by  the solution of the time-dependent Schr\"{o}dinger equation with multiple
coupled states using a
wave packet method.
A diabatization procedure is
proposed to obtain the electronic couplings between quasidiabatic potentials of $^1\Sigma^+$ symmetry for HF.
By including these couplings between the neutral states, the cross section for ion-pair formation increases with 
about two orders of magnitude compared with the cross section for direct dissociation.
Qualitative agreement with the measured cross section is obtained. The oscillations in the calculated cross section
are analyzed. 
The cross section for ion-pair formation in electron recombination with DF$^+$ is calculated to 
determine the effect of isotopic substitution.
\end{abstract}

\pacs{Valid PACS appear here}
\maketitle

\section{Introduction}
In low-temperature plasmas there is a significant concentration of molecular ions. Processes such as dissociative recombination (DR) and resonant ion-pair (RIP) formation, that modify the charge 
and energy balance in these plasmas, are of great importance. To model the plasma environments correctly, a deeper understanding of the mechanisms and cross sections of these processes is needed. 

In DR, an electron is captured by the molecular ion losing its energy either to electronic or to ro-vibronic excitations of the resulting neutral molecule, corresponding to the ``direct''  or ``indirect'' processes respectively of DR as proposed by Bardseley~\cite{bardsley68}. The molecule then stabilizes by dissociating into fragments. In the DR process, the fragments are neutral species, whereas in the RIP process the fragments consist of an ion-pair. 

The cross section of ion-pair formation in electron recombination have been measured for a limited number of ions such as HD$^+$~\cite{zong99,larson00,neau02}, H$_3^+$~\cite{peart79,yousif93,kalhori04}, NO$^+$~\cite{lepadellec01}, OH$^+$~\cite{larson00} and HF$^+$~\cite{djuric01,wolf09}. Several of these measurements used the ion-storage ring technique~\cite{zong99,larson00,neau02,kalhori04,lepadellec01,djuric01,wolf09}. The advantage of using storage rings is that the ion has time to relax into its vibrational ground state prior to the experiment, so there is a well defined initial state. Also, the ionic fragments can be separated from the neutral fragments in the bending magnets, and hence the final ion-pair channel is also well defined. Thus, theoretical studies on the ion-pair process can be compared to experimental studies and give more insight into the physics of the reaction.  

So far only a few theoretical studies on resonant ion-pair formation have been presented. Only the systems HD$^+$~\cite{larson00,larson01}, HeH$^+$~\cite{larson99}, and H$_3^+$~\cite{kalhori04,larson06,roos07} have been examined. In these studies semi-classical methods~\cite{larson00,larson06} or wave packets~\cite{larson01,larson99,kalhori04,roos07} have been used to describe the dynamics. The modeling of the RIP process is a theoretical challenge. Assuming, the direct process, a resonant state is created when the electron is captured by the molecular ion. Both the resonant state potentials and their corresponding autoionization widths must be well described. When the potential of the resonant state has crossed the ionic ground state potential, autoionization is no longer possible and the resonant state becomes electronically stable. 
It will then interact with the manifold of Rydberg states with potentials situated below the ion. To be able to capture the true dynamics of the RIP process, accurate potential energy curves, autoionization widths and electronic couplings need to be calculated over a large range of internuclear distances. The nuclear dynamics have to be explored from the Franck-Condon region where the electron is captured and
into the asymptotic region. The
system 
may take different pathways on the way to dissociation into the ion-pair channel. Several interesting quantum effects may be studied, such as interferences due to the separate routes to the ion-pair limit yielding (St\"{u}ckelberg) oscillations~\cite{stuckelberg32} in the cross section for the reaction or tunneling through barriers in the potentials yielding resonant structures (shape resonances) in the cross section. 

In this paper we present calculations on the ion-pair formation in  electron recombination with HF$^+$, i.e., the process:
\begin{equation}
\mbox{HF}^{+} + \mbox{e}^{-} \longrightarrow \mbox{H}^{+} + \mbox{F}^{-}.
\end{equation}

The cross section for this reaction has been measured~\cite{djuric01} for collision energies ranging from 0.0001 to 1 eV, using the ion-storage ring CRYRING, where the F$^-$ fragments were detected. Since the electron affinity of F is large, this ion-pair limit has a threshold energy of only 0.017 eV relative to the $v=0$ level of the ion. This is within the rotational energy spread of the HF$^+$ target ion and therefore the measured cross section did not reveal any threshold effects. The ion-pair cross section was found to be relatively large, about $14$ \% of the DR cross section at a collision energy of 0.02 eV. The cross section for RIP shows interesting structures. Using photo ion-pair experiments, where the dissociation into the ion-pair H$^+$ + F$^-$ is studied using photon excitation of the ground
state $X^1\Sigma^+$ HF molecule into one of the
bound Rydberg states, similar structures have been observed ~\cite{yencha95,hu05,hu06}.
Very recently, this RIP process has also been studied experimentally using the TSR ion-storage ring~\cite{wolf09}. In this experiment, the absolute cross section was not measured. However, the shape and the structures in the relative cross section were very similar to those measured in CRYRING.
Using an imaging technique where the neutral H($n=2$)+F($^2P_{3/2}$) fragments were detected, the rotational distribution as well as the population of the HF$^+$($X^2\Pi_{3/2}$) and HF$^+$($X^2\Pi_{1/2}$) components were explored~\cite{wolf09}. 

We have previously studied the direct mechanism of DR of HF$^+$~\cite{roos08}. We calculated 30 resonant states of HF and a time-independent method was used to determine the total cross section for dissociative recombination.
Autoionization from the resonant states was included in the model.
The electronic couplings between the neutral states were neglected. It was thus assumed that the flux captured into a resonant state would follow that state diabatically out into the asymptotic region, without a redistribution of the flux.
This calculation produced a total cross section for DR of HF$^+$ in qualitative agreement with the measured cross section.
Sharp threshold effects were seen where the asymptotic states become energetically open. The measured cross section below 0.04 eV was not reproduced and this was believed to be due to the neglect of the electronic couplings between the neutral states. The lowest $^1\Sigma^+$ resonant state, in the quasidiabatic representation, dissociates into the ion-pair limit,
H$^+$ + F$^-$. The calculated cross section for the direct dissociation into the ion-pair state was two orders of magnitude lower than the experimental results from CRYRING~\cite{djuric01}. In the previous study, we
presented the hypothesis that inclusion of the electronic couplings in the model might increase the cross section of ion-pair formation. Flux captured by higher resonant states could couple into bound Rydberg states that then predissociate by the electronic couplings with the ion-pair state.

In the present study, we propose a diabatization procedure that allows us to obtain not only the quasidiabatic potentials, but also the electronic couplings among the neutral states. This method is applied to diabatize the $^1\Sigma^+$ states of HF relevant in dissociative recombination and especially ion-pair formation. The cross section is then studied by propagating wave packets on coupled states. Section \ref{sec:pec} describes the calculation of relevant potentials and autoionization widths.
The diabatization procedure is outlined in section \ref{sec:diab}, while section \ref{sec:dyn} describes the treatment of the reaction dynamics. The results are presented in section \ref{sec:res}. Atomic units are used throughout the paper unless otherwise stated.

\section{\label{sec:pec}Potential Energy curves and autoionization widths}
In our earlier study on HF~\cite{roos08}, potential energy curves and autoionization widths were calculated {\it ab initio} by combining multi reference configuration interaction (MRCI) structure calculations with electron scattering calculations. In the present study, we are only interested in the $^1\Sigma^+$ symmetry containing the ion-pair state. The potentials and autoionization widths of the resonant states situated above the ionic ground state potential are taken from the earlier scattering calculations. However, electronically bound states are recalculated using the MRCI method on a much finer grid for internuclear distances ranging from $1.0$ a$_0$ to $9.0$ a$_0$. This is done in order to accurately resolve the avoided crossings among the adiabatic states induced by the couplings between resonant and Rydberg states.
For computational details we refer the reader to the previous study~\cite{roos08}. 
We only present a brief summary of the calculation method.
For the MRCI, the molecular orbitals are first determined using a SCF calculation followed by a MRCI calculation on the neutral ground state to calculate natural orbitals and hence a more compact representation of the orbitals is obtained. These natural orbitals are then used in order to carry out the MRCI calculations on ionic ground state ($X^2\Pi$) and the excited states of HF of $^1\Sigma^+$ symmetry. The MRCI calculation has reference configurations created by all excitations among the seven natural orbitals with highest occupations (3 $\sigma$-orbitals and 4 $\pi$-orbitals), except for the lowest ($1\sigma$) core orbital that is kept doubly occupied in all reference configurations. Single excitations from this set of reference configurations into the virtual orbitals are included. 
At each internuclear distance, 25 roots are calculated.

The same target wave function is used in the electron scattering calculations carried out with the complex Kohn variational method~\cite{rescigno95}. By fitting the eigenphase sum of the transition matrix to a Breit-Wigner form~\cite{geltman97} both the autoionization with $\Gamma(R)$ and the resonance energy $E_{res}(R)$ can be determined. The potential energy curve of the resonant state is obtained by adding the ionic potential to the resonance energy. In the present calculation, three resonant states of $^1\Sigma^+$ symmetry are included.

\section{\label{sec:diab} Diabatization procedure}
Rydberg states have the same character as the ground state of the ion $\left[(1\sigma)^2(2\sigma)^2(3\sigma)^2(1\pi)^3\right]$ plus an outer electron in a diffuse orbital. The resonant states are Rydberg states that converge to electronically excited ionic cores. These cores all have the $(3\sigma)$ orbital singly excited. By following the configurations of the resonant states we obtain an ``initial guess" of the quasidiabatic potential curves. This was the method applied to obtain the uncoupled diabatic potentials in our previous study~\cite{roos08}. As mentioned above, the cross section for ion-pair formation, calculated by only including the resonant state that diabatically correlates with
the ion-pair limit, was two orders lower in magnitude than what has been measured~\cite{djuric01}. In order to improve this calculation, the electronic couplings between the neutral states must be included. Here, we propose a method where the adiabatic potentials and the ``initial guess" of the quasi diabatic potentials are used in
order to determine not only the quasidiabatic potentials but also the electronic couplings. 

\subsection{Extended two-by-two transformation}
Assume two adiabatic states interact in the vicinity of an avoided crossing. 
In the two-state model, the adiabatic potential matrix {\bf V} with matrix elements $V_{ij}=V_i(R)\delta_{ij}$ can be a transformed into a diabatic potential matrix {\bf U} using the transformation matrix {\bf P}. The diagonal elements of {\bf U} are the diabatic potentials and the off-diagonal elements correspond to the electronic coupling. The transformation is given by
\begin{eqnarray}
\mathrm{\mathbf{U=P}^{-1}\mathbf{VP}},
\end{eqnarray}
where the unitary, orthogonal transformation matrix, in the two-state problem can be written as
\begin{eqnarray}
\mathrm{\mathbf{P}}=\left(\begin{array}{c c}
\cos[\gamma(R)] & -\sin[\gamma(R)]\\
\sin[\gamma(R)] & \cos[\gamma(R)]
\end{array}\right).
\end{eqnarray}
Asymptotically, far from the avoided crossing, we assume that the adiabatic and diabatic potential curves are identical. 
This assumption forces the rotational angle $\gamma(R)$ to be a function that goes from $0$ to $1$. 

The two state model is now extended to include more states by treating each crossing as a two-by-two transformation $\mathbf{P}_i$, a rotation where only two states are included and the rest are left unchanged. If there are $n$ crossings, the diabatic potential matrix, $\mathbf{U}$, is given by
\begin{eqnarray}
\label{eq:utrans}
\mathrm{\mathbf{U}=\mathbf{P}_n^{-1}\ldots\mathbf{P}_2^{-1}\mathbf{P}_1^{-1}\mathbf{V}\mathbf{P}_1\mathbf{P}_2\ldots\mathbf{P}_n},
\end{eqnarray}
where the matrices $\mathbf{P}_i$ are of the form (in the case of rotation among state 1 and 2)
\begin{eqnarray}
\mathrm{\mathbf{P}_i}=\left(\begin{array}{c c c c c}
\cos(\gamma_i) & -\sin(\gamma_i) & 0 & 0 & \cdots\\
\sin(\gamma_i) & \cos(\gamma_i) & 0 & 0 & \cdots\\
0 & 0 & 1 & 0 & \cdots\\
0 & 0 & 0 & 1 & \cdots\\
\vdots & \vdots & \vdots & \vdots & \ddots\\
\end{array}\right).
\end{eqnarray}
The rotational angles are assumed to have the following analytical form
\begin{eqnarray}
\gamma_i(R)=\frac{\pi}{4}[1+\tanh(\alpha_i(R-R_i))].
\end{eqnarray}
An analytical expression for the total transformation matrix is now obtained and this matrix depends upon the unknown parameters of the rotational angles. The total transformation matrix
can be shown to be an orthogonal matrix.
After the diabatic potential matrix, $\mathbf U$, is set up we perform an optimization procedure, where the unknown parameters $\alpha_i$ and $R_i$ of the rotational angles are determined in order to optimize the agreement between the diagonal elements of $\mathbf{U}$ and the estimated diabatic potentials obtained by tracking the configurations as described above. When the parameters are optimized, also the electronic couplings are obtained as the off-diagonal elements of the matrix $\mathbf{U}$.
By construction [see equation (\ref{eq:utrans})], the obtained quasidiabatic potentials have the adiabatic potentials as eigenvalues. 
However, we do not show that the couplings elements of the nuclear kinetic energy operator (non-adiabatic interactions) for the quasidiabatic states are identically zero. Therefore, the proposed method do not produce any strict diabatic states~\cite{maed82}, but rather quasidiabatic potentials and couplings.

\begin{figure*}
\includegraphics*[54,344][569,497]{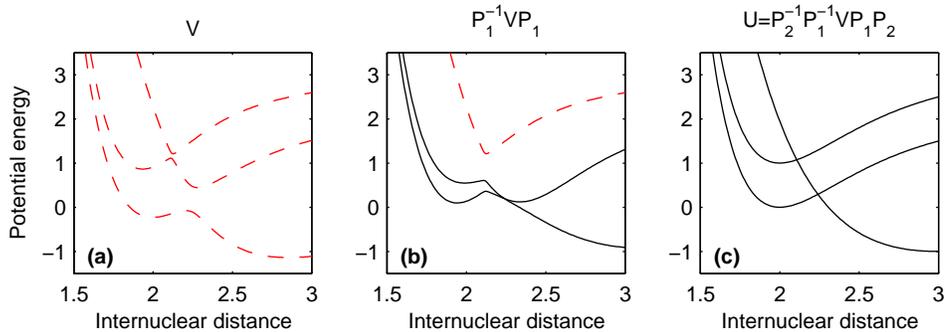}
\caption{\label{fig:Model} (Color online) Schematic picture of the extended transformation. In (a), the three adiabatic potentials are shown. In (b) state 1 and 2 are rotated and finally in (c) a second rotation is performed among state 3 and 2.}
\end{figure*}

Figure~\ref{fig:Model} shows a schematic picture of this transformation in the case of three states and two crossings.

\subsection{Quasidiabatic potential transformation}
The diabatization procedure described above is now used to obtain the potential energy curves of HF of $^1\Sigma^+$ symmetry. The experimental cross section for ion-pair formation is measured for collision energies up to 1 eV~\cite{djuric01}. Therefore, only the three lowest resonant states are included in the diabatization procedure. More energy is needed in order to capture the electron and form the higher resonant states. Furthermore, from the structure calculations the ground state of HF and five Rydberg states are obtained. 

\subsubsection{Model I: 8 coupled states}
In the first model, the ground state of HF ($X^1\Sigma^+$, labeled here with $U_{11}$) is excluded in the diabatization procedure. The ground state is assumed not to couple to the excited states of HF. Hence, in model I eight states are included in the diabatization procedure, five quasidiabatic Rydberg states [labeled with $U_{22}, \ldots U_{66}$] and three resonant states [$U_{77},U_{88}$ and $U_{99}$]. State $U_{77}$ corresponds to the resonant state associated with the ion-pair limit. The optimization procedure is performed for internuclear distances $1.0$ a$_0\leq R \leq 9.0$ a$_0$. 
  
\begin{figure}
\rotatebox{0.0}{\includegraphics[width=1.0\columnwidth]{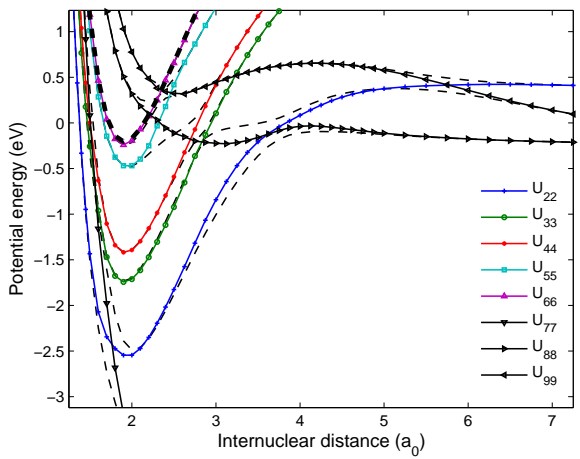}}
\caption{\label{fig:potI} (Color online) \textbf{Model I}. Adiabatic (dashed) and quasidiabatic potential curves [solid (color online) lines] in the region of avoided crossings. Also the ground state of the HF$^+$ ion is displayed with the thick dotted curve. The energy scale is relative to the ground vibrational state of the ion.}
\end{figure}

In Figure~\ref{fig:potI}, the adiabatic (dashed) and quasidiabatic (solid) potential curves of HF are shown as well as potential (thick dashed) of the HF$^+$ ion. The potential curves of the quasidiabatic resonant states are black online.

\begin{figure}
\rotatebox{0.0}{\includegraphics[width=0.95\columnwidth]{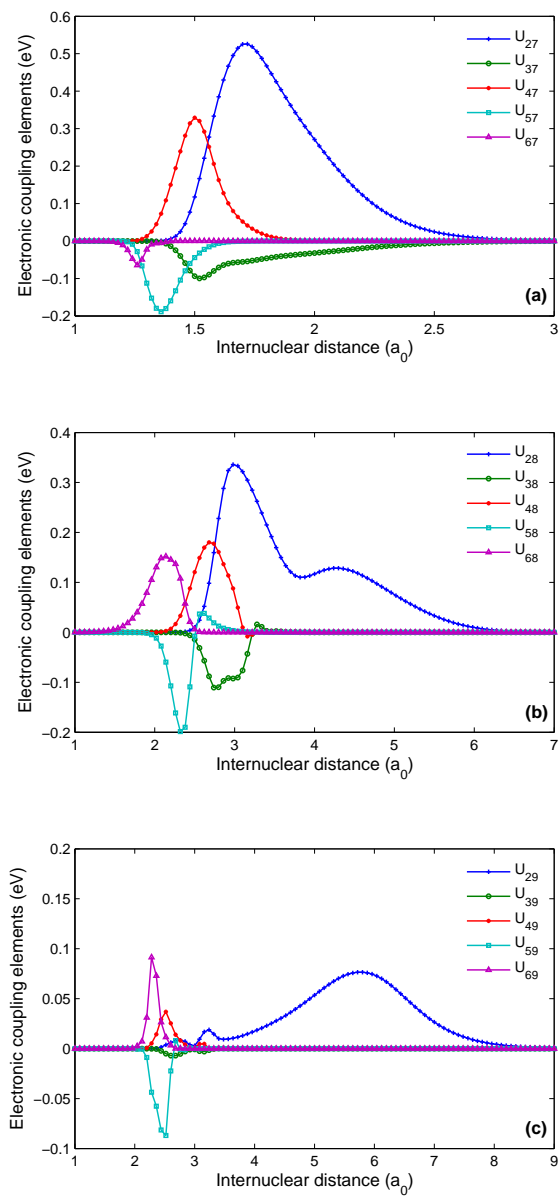}}
\caption{\label{fig:CoupI} (Color online) \textbf{Model I}. Electronic couplings between Rydberg states and the resonant states (a) $U_{77}$, (b) $U_{88}$, and (c) $U_{99}$.}
\end{figure}

The calculated electronic couplings between the neutral states of HF are displayed in Figure~\ref{fig:CoupI}. The electronic couplings between the Rydberg states and the resonant states [$U_{77}$ in (a), $U_{88}$ in (b), and $U_{99}$ in (c)] are shown.
Note that the electronic couplings are localized to the regions of the avoided crossings between the adiabatic states.

The potential curves and autoionization widths are extrapolated to larger and smaller internuclear distances. 
The extrapolation toward smaller distances, is carried out as described in our previous study~\cite{roos08}.
The asymptotic limits of the potential energy curves have to be determined. Here the experimental energies of the asymptotic limits~\cite{NIST:AtSpecData} are used. Since the spin-orbit coupling in the system is neglected, the mean values of the energy levels with different $J$ for the F atom and mean values for states with equal principal quantum number $n$ for the H atom are used for the calculation of the asymptotic limits. Wigner-Witmer rules~\cite{herzberg50} are then applied to determine the number of states of $^1\Sigma^+$ symmetry that goes to each asymptotic limit. In table I, the asymptotic limits for states included in the present study are listed.
The potentials that are associated with neutral fragments are assumed to have reached their asymptotic limits at $R=50.0$ a$_0$. The ion-pair state has an asymptotic Coulomb form
\begin{eqnarray}
V(R)=V_{final}-\frac{1}{R}-\frac{\alpha}{2R^4},\label{asyform}
\end{eqnarray}
where $V_{final}$ is the asymptotic energy limit and
$\alpha=17.581$ a$_0^3$ is the polarizability of F$^-$~\cite{fritsch80}. We assume that the ion-pair state has this form at internuclear distances $R\geq 20$ a$_0$ and interpolate between the calculated ion-pair state and the asymptotic form of the ion-pair using spline interpolation. 
The electronic couplings will go smoothly to zero outside the region where the potentials are diabatized.

\begin{table}
\caption{\label{tab:asy} Asymptotic energy limits of the resonant states included in the present study. The energies are relative to the $v=0$ vibrational level of HF$^+$.}
\begin{ruledtabular}
\begin{tabular}{cccc}
 &Atomic States &Molecular States &Energy (eV)\\
\hline
\\
1& H($n=1$) + F($^2P$) & $U_{77}\ {\rm(Model\ II)}$ & $-10.1552$ \\
2& H$^+$ + F$^-$ & $ U_{77}\ {\rm(Model\ I)},\ U_{11}\ {\rm (Model\ II)}$ & $0.0170$ \\
3& H($n=2$) + F($^2P$) & $U_{22},\ U_{88},\ U_{99}$ & $0.0437$ \\
4& H($n=4$) + F($^2P$) & $U_{33},\ U_{44}$ & $2.5934$ \\
5& H($n=1$) + F($^2P$) & $U_{55},\ U_{66}$ & $2.8993$ \\
\end{tabular}
\end{ruledtabular}
\end{table}

\subsubsection{Model II: 9 coupled states}
\begin{figure}
\rotatebox{0.0}{\includegraphics[width=1.0\columnwidth]{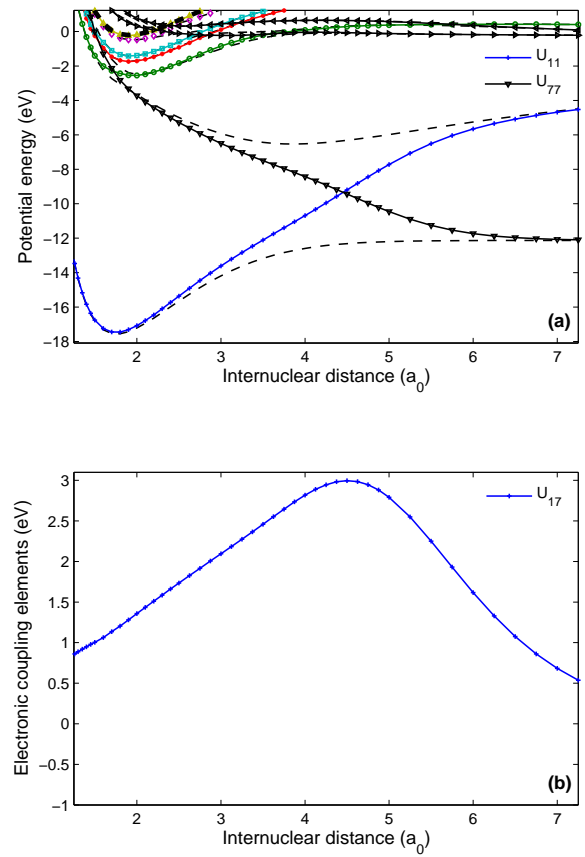}}
\caption{\label{fig:potCoupII} (Color online) \textbf{Model II}. Figure (a) shows the adiabatic potentials (dashed lines) and the quasidiabatic potentials (solid lines) of HF in the region of avoided crossings. Also the ground state of HF$^+$ is displayed with the thick dashed curve. 
Figure (b) shows the electronic coupling between the ground state and ion-pair state.}
\end{figure}

In the literature, there has been a discussion about a possible change of character of the ground state of HF when the internuclear distance is stretched ~\cite{bettendorff82,chaudhuri01}. At small distances, the ground state has an ion-pair character, while as larger distances it goes covalently to the lowest limit H($n=1$)+F($^2P$). Hence, there is a 
large avoided crossing between the ground state of HF and the ion-pair state.
Similar avoided crossings are possessed by the 
alkali halides such as  LiF, LiCl and NaI~\cite{werner81,weck04,alekseyev00}.
In model II, we examine the influence of this avoided crossing upon the ion-pair formation.
Therefore, the lowest quasidiabatic potential ($U_{11}$) is in model II associated with the ion-pair limit, while the lowest resonant state ($U_{77}$) goes to the ground asymptotic fragments (see table~\ref{tab:asy}). Nine states are then diabatized as described above. In Figure~\ref{fig:potCoupII} (a) the resulting adiabatic and quasidiabatic potential curves are displayed. In (b) the electronic coupling between the ground state and ion-pair state is shown. The electronic couplings between the higher excited states of Model II are similar to the corresponding couplings in Model I.
As can be seen in Figure~\ref{fig:potCoupII} (b), the coupling element between the states with $U_{11}$ and $U_{77}$ potentials is an order of magnitude larger than the other coupling elements. The size of this coupling element is proportional to the energy difference between the adiabatic curves at the avoided crossing between these states. This large electronic coupling indicates that the nuclear dynamics most probably will go adiabatically in this region.

\section{\label{sec:dyn} Reaction Dynamics with wave packets}
The cross section for dissociating into the different channels, particularly into the ion-pair state, is calculated by propagating wave packets on the coupled neutral states. The electron capture will initiate wave packets on the resonant states. 
For resonant state $i$ (with $i\geq7$ in model I and II mentioned above), we have~\cite{mccurdy83}
\begin{equation}
\Psi_i(t=0,R)=\sqrt{\frac{\Gamma_i(R)}{2\pi}}\chi_{v=0}(R),
\end{equation}
where $\Gamma_i$ is the autoionization width for the resonant state and $\chi_{v=0}$ is the $v=0$ vibrational wave function of the ionic ground state. 

It is here assumed that the electron is captured by the molecular ion in its lowest vibrational state. This is adequate when we compare with measurements using the ion-storage ring where the ion vibrationally relax prior to the experiment~\cite{djuric01}. However, the ions in the storage ring will be rotationally excited. 
Furthermore, the spin-orbit coupling is excluded in the calculation of the relevant potentials. 
The fine-structure splitting of the $X ^2\Pi_{3/2}$ and $X^2\Pi_{1/2}$ components of HF$^+$ is about 30 meV~\cite{gewutz75} and both of these components will exist in the storage ring experiments~\cite{wolf09}.

The wave packets are initiated only on the resonant states. The non-adiabatic coupling between the ionization continuum and the Rydberg states is thus neglected and therefore, only the direct mechanism of DR is studied. 

The nuclear dynamics is explored by solving the time-dependent Schr\"odinger equation
\begin{equation}
\label{eq:se}
i\frac{\partial}{\partial t}{\bf \Psi}(t,R)=-\frac{1}{2\mu}{\bf I}\frac{\partial^2}{\partial R^2}{\bf \Psi}(t,R) + 
{\bf U}(R){\bf \Psi}(t,R)
\end{equation}
numerically by using a Cranck-Nicholson propagator~\cite{goldberg67}. In this study we have propagated wave packets on both coupled and uncoupled potentials. In the uncoupled case ${\bf U}$ is a diagonal matrix containing only the quasidiabatic potentials of the resonant states. Autoionization from the resonant states are included by letting these potentials be complex when their potential curves are situated above the ionic potential curve. In the local approximation for treating autoionization, also called the ``Boomerang model''~\cite{mccurdy83}, the resonant state potentials are given by 
\begin{equation}
\tilde{U}_{ii}(R)=U_{ii}(R)-i\frac{\Gamma_i(R)}{2}.
\end{equation}
In our previous study of HF~\cite{roos08} we examined the validity of this approximation, and indeed it is valid for the states included in the present study.

When the wave packets have been propagated out into the asymptotic region they are projected onto energy normalized eigenstates of the fragments $\left[\Phi^i_E(R)\right]$ and the cross section can be calculated as~\cite{larson01}
\begin{equation}
\sigma_i(E)=\frac{2\pi^3}{E}g\left|\langle\Phi^i_E(R)|\Psi_i(t_{\infty},R)\rangle\right|^2,
\end{equation}
where $g$ is the multiplicity ratio of the neutral state and the ionization continuum.
In the present study a grid ranging from $R=0.8\ {\rm a_0}$ to $R=300\ {\rm a_0}$ with grid steps of $dR=0.01\ {\rm a_0}$ is used. The wave packets are propagated with a time step of $dt=0.1\ {\rm a.u.}$. For the uncoupled potentials the final propagation time is $t_{final}=1000\ {\rm a.u.}$, while $t_{final}=4000\ {\rm a.u.}$ is needed for the coupled systems. In the coupled state calculations, part of the wave packets are trapped in bound Rydberg states before they predissociate and therefore a longer propagation time is needed to reach convergence.

\section{\label{sec:res}Results and discussion}
The wave packets are now propagated on the coupled potentials of $^1\Sigma^+$ symmetry where three resonant states are included. As mentioned above, model I includes five Rydberg like potentials and model II also includes the ground state of HF and the electronic coupling between the ground state and the ion-pair state.
To test the role of the electronic couplings, we also propagate the wave packets on uncoupled potentials.
Furthermore, the roles of rotation as well as the spin-orbit splitting of the target ion state are addressed. 
Finally, the isotope effect on ion-pair formation is studied by calculating the cross section for the DF$^+$ ion. 

\subsection{Coupled system Model I}
\begin{figure}
\rotatebox{0.0}{\includegraphics[width=1.0\columnwidth]{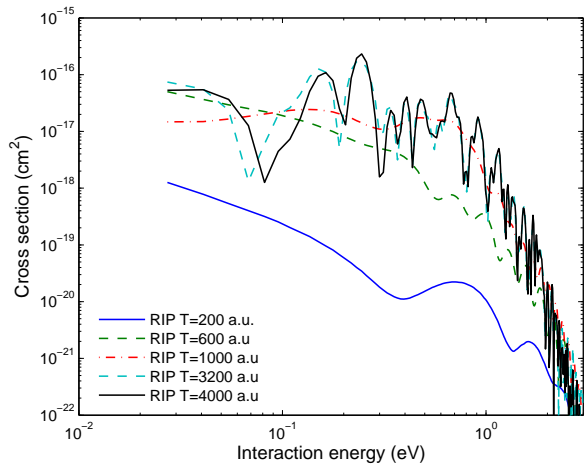}}
\caption{\label{fig:TimeDep} (Color online) Time dependence of the ion-pair formation cross section calculated using the 8 couples states of model I. At t=4000 a.u, the cross section has converged.}
\end{figure}

When the wave packets are propagated on the coupled potentials, parts of the wave packets that are initially captured into the resonant states will couple into the bound Rydberg states and temporary get trapped in these states before predissociation is induced by the electronic coupling between the bound state and states that are open for dissociation. 
Feshbach resonances are formed that creates oscillatory structures in the cross section 
for ion-pair formation. In order to obtain convergence of these structures,
the  propagation time must be long enough for the wave packets to form the temporary trapped resonant state and subsequently predissociate. 
We have found however that after a time of $t=3200\ {\rm a.u.}$ the structures in the cross section do not change significantly.
Therefore at a final propagation time of $t_{final}=4000\ {\rm a.u.}$ the calculation is converged.
 In order to avoid problems with reflection against the end of the grid or against the absorbing potentials for some of the states a much larger grid than in the previous calculation ~\cite{roos08} was used.
  Figure~\ref{fig:TimeDep} shows the ion-pair cross section calculated for different propagation times, where all 8 coupled states in model I are included. 
For a propagation time of $t=200\ {\rm a.u.}$ only the direct dissociation along the ion-pair state is seen. This cross section underestimates the measurements with about two orders of magnitude. At $t=600\ {\rm a.u.}$ on the other hand, flux from the higher resonant states has predissociated through the Rydberg states into the ion-pair state causing an increase in the cross section.
However, the calculation at $t=600\ {\rm a.u.}$ is not converged with respect to time. We can see that while a time of $t=1000\ {\rm a.u.}$ is enough to obtain the amplitude of the cross section, at later times, structures appear and as mentioned above, these structures do not change significantly after a propagation time of $t=3200\ {\rm a.u.}$. Longer time is needed in order to converge the structures at lower energies. The final propagation time is set to $t_{final}=4000\ {\rm a.u.}$. Such behaviour has been observed previously~\cite{morisset07}.

\begin{figure}
\rotatebox{0.0}{\includegraphics[width=1.0\columnwidth]{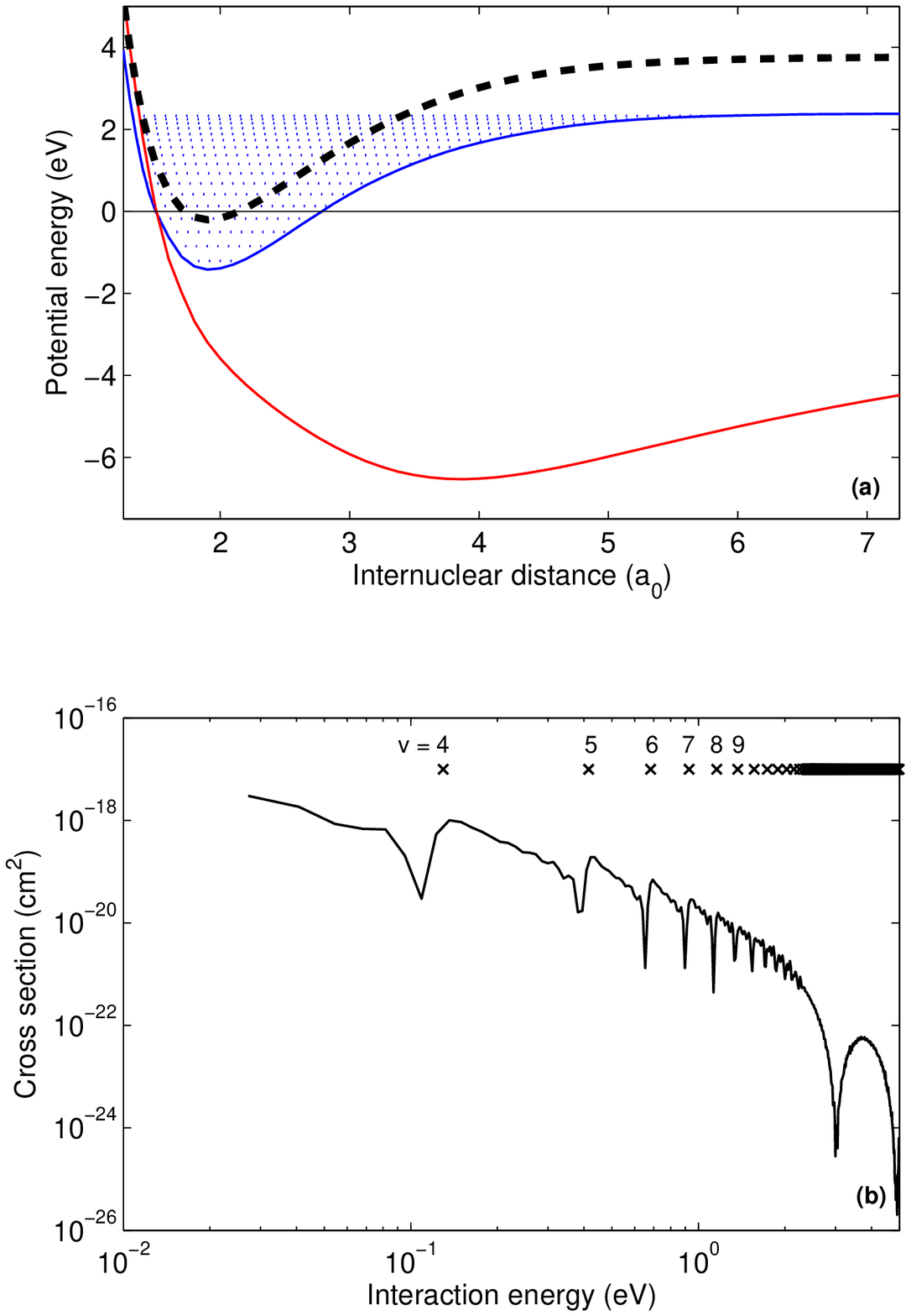}}
\caption{\label{fig:Res1Ryd3} (Color online) In (a), the potentials included in the test calculation are displayed. These are the ion-pair state $U_{77}$ (red online) and the bound Rydberg state $U_{44}$ (blue online). The energies of the vibrational levels of the Rydberg state are displayed with the dotted (blue online) lines. The potential of the ion is the thick dashed curve.
In (b) the cross section for ion-pair formation obtained using the test calculation is shown. The structures in the cross section can be explained by temporary formation of the bound vibrational levels of the Rydberg state. The energies of these vibrational levels are marked with crosses.}
\end{figure}

In order to understand the oscillatory structures found in the ion-pair cross section shown in Figure~\ref{fig:TimeDep}, a test calculation is carried out where only the ion-pair state ($U_{77}$) and the third Rydberg state ($U_{44}$) as well as the couplings between these states are included. These potentials are displayed in Figure~\ref{fig:Res1Ryd3} (a) where also the ion potential (thick dashed) curve is shown.
The dotted lines are the vibrational energy levels of the Rydberg state. 
The calculated cross section for ion-pair formation [see Figure~\ref{fig:Res1Ryd3} (b)] shows sharp dips and the positions of these dips match perfectly with the energy eigenvalues of the vibrational levels of the Rydberg state, marked with crosses in the figure.
Note that this kind of structure disappears at the energies larger than the dissociation limit of the Rydberg state. 
For this test calculation where only two coupled states are included, the alternative time-independent approach used in our previous study~\cite{roos08} is also applied in order to check the convergence of the wave packet propagation. A very good agreement for the cross section calculated using the two different methods is found. 

\begin{figure}
\rotatebox{0.0}{\includegraphics[width=1.0\columnwidth]{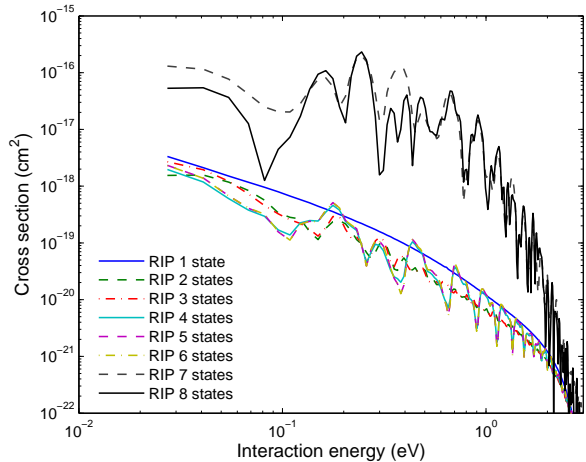}}
\caption{\label{fig:CoupledI} (Color online) Cross section for the ion-pair state when different number of states are included in the coupled system}
\end{figure}

In model I, five Rydberg states are included and each will contribute with these kinds of oscillations in more or less overlapping energy intervals causing the rich structure seen in the ion-pair cross section for the coupled system. 
In order to study the effects on the ion-pair cross section from the couplings between the neutral states, one state at the time is added to the model.
The resulting ion-pair cross sections are shown in Figure~\ref{fig:CoupledI}.
The first calculation only contains the ion-pair state and as seen before, the cross section is underestimated by about two orders of magnitude.
Then, the bound Rydberg states ($U_{22},\ldots U_{66}$) are added one after the other and this causes the dips in the cross section mentioned above.
Note that the magnitude of the cross section is not dramatically affected. When the first Rydberg states is added, the cross section decreases as expected and some structures emerge. Adding the second Rydberg state only affects
the cross section slightly. This is expected since the magnitude of the electronic coupling between the ion-pair state and the second Rydberg state is about one order of magnitude smaller than the electronic coupling between the ion-pair state and the first Rydberg state. When the third Rydberg state is added, more pronounced structures emerge.
Adding the fourth and fifth Rydberg state does not effect the cross section significantly since the couplings to these states are relative small.
When the second resonant state ($U_{88}$) is added to the model, the cross section of the ion-pair is significantly increased since flux is predissociating from this state through the Rydberg states and into the ion-pair state. This can be seen by the black dashed curve for the cross section in figure~\ref{fig:CoupledI}. It should be noted that at low collision energies, the second resonant states has the largest electron capture probability at low collision energies. Also the sizable coupling elements between the second resonant states and the Rydberg states explains the increase of the ion-pair cross section when this state is added.
When the third resonant state is added, the cross section do not change significantly, except for the energy interval between $0.3-0.4\ {\rm eV}$. This is the energy window where the crossings between this state and the Rydberg states occur, see Figure~\ref{fig:potI}. 

In the present study, the indirect mechanism of dissociative recombination is neglected. We thus neglect the non-adiabatic coupling between the Rydberg state and the ionization continuum. Induced by this coupling, the electron could be directly captured into a ro-vibronically excited Rydberg state that then predissociate by the electronic coupling to the ion-pair state. 
This could in principle affect the cross section for ion-pair formation. However, it is shown that the indirect mechanism is most important at very low collision energies, typically below 0.1 eV~\cite{bates91,florescu03,ngassam06}. We believe the inclusion of the indirect mechanism could affect the cross section at very low collision energies and furthermore it would cause sharp structures in the cross section. These structures are not resolved in the measured cross section using the storage ring experiment~\cite{djuric01}. The indirect mechanism is best adressed using Multi-Channel Quantum Defect Theory (MQDT)~\cite{florescu-mitchell06}. However, standard MQDT calculations do not give final state distributions and hence could not obtain the ion-pair cross section. The indirect mechnaism has been included in wave packet studies~\cite{larson05}. This yielded accurate branching ratios, but the magnitudes of the partial cross sections were difficult to converge.

\begin{figure}
\rotatebox{0.0}{\includegraphics[width=1.0\columnwidth]{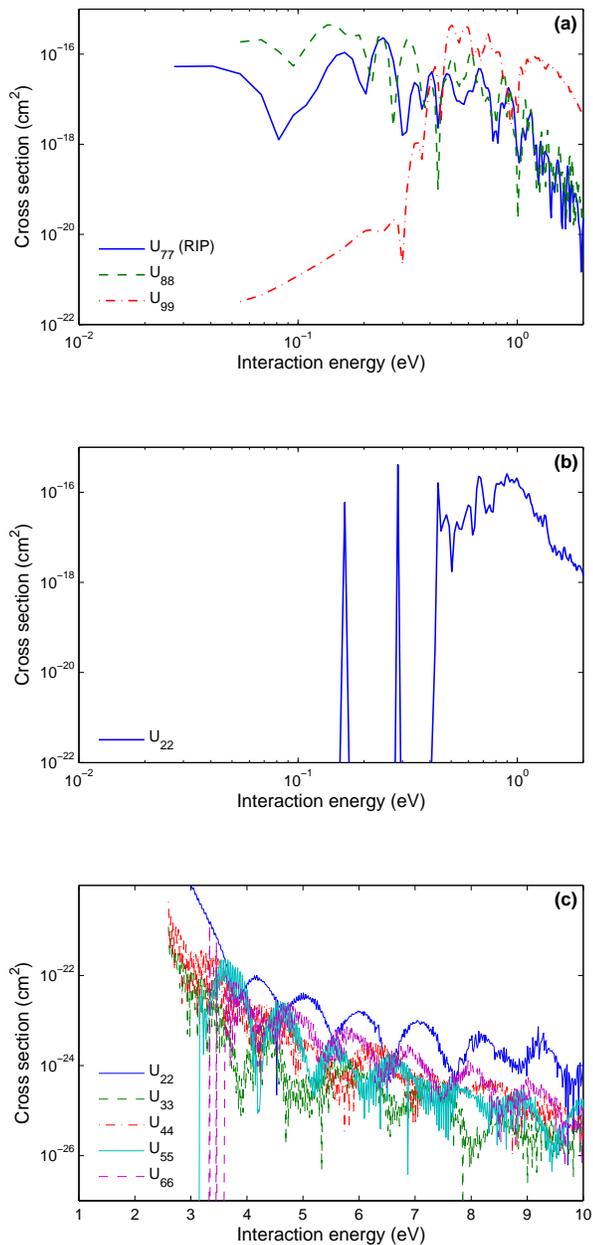}}
\caption{\label{fig:Partial8} (Color online) Partial cross section for the 8 states included in model I, in (a) the cross sections for the resonant states are displayed, while (b) and (c) shows the cross section for dissociation along the Rydberg states.}
\end{figure}

The partial cross sections for all states included in model I are shown in Figure~\ref{fig:Partial8}. In (a), the cross sections for the resonant states are displayed. In (b) the cross section for first Rydberg state is shown, while in (c) also the cross sections for the higher Rydberg states are displayed.
Only the lowest Rydberg states and the two resonant states are energetically open at lower energies (see Table I). 
The lowest Rydberg state has a barrier in its potential which explains the sharp onset around 0.4 eV. The resonances below (and above) the barrier are so-called tunneling shape resonances. Similar shape resonances have been found in for example the cross section for dissociative recombination of HeH$^+$~\cite{larson05}.

\subsection{Model I vs. Model II}
\begin{figure}
\rotatebox{0.0}{\includegraphics[width=1.0\columnwidth]{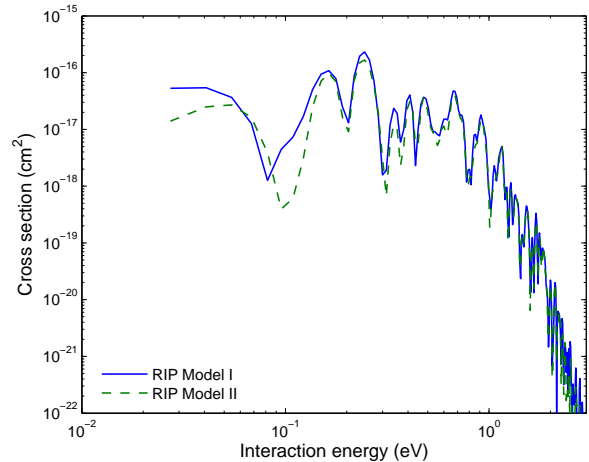}}
\caption{\label{fig:ModI_ModII} (Color online) Cross section for ion-pair formation calculated using model I and model II of potential curves and couplings.}
\end{figure}

In model II, the crossing of the ground state and ion-pair state are included and we investigate the importance of such a crossing on the dynamics. The cross section for ion-pair formation, calculated using the two different models is displayed in Figure~\ref{fig:ModI_ModII}. Note that the inclusion of the curve crossing does not affect the magnitude of the cross section. The large electronic coupling between the ground state and ion-pair state (see Figure~\ref{fig:potCoupII}) forces the wave packet to propagate adiabatically through the region of the crossing.
At the higher energies, model II gives a smoother cross section. In model I, part of the wave packet will oscillate back and forth around the minimum ion-pair potential and hence cause interferences that 
require a longer propagation time for convergence. In model II however, the possibility of dissociating
into the ground state limit also exists, and less flux will oscillate and interfere.
This results in a smoother cross section.
 At the very low energies, the lack of total convergence with respect to time is more noticeable, and we believe that this is the main reason
 for the discrepancy in the cross section between the two models in that region. 

\subsection{Effects from spin orbit splitting and rotation}
In the structure calculation, the spin orbit coupling is neglected and therefore we do not see the splitting of the ground state for the ion into its two components $X^2\Pi_{3/2}$ and $X^2\Pi_{1/2}$. The splitting of these components ias of the order of 30 meV ~\cite{fritsch80}. In the experiments~\cite{djuric01,wolf09}, both spin-orbit components of the ion will be populated with some unknown distribution. The most recent experiment carried out using the TSR storage ring~\cite{wolf09}, shows fragments produced in dissociative recombination with the HF$^+$ ion in the $X^2\Pi_{3/2}$ state with the rotationally excited $J=7/2, 9/2, \ldots, 15/2$ levels in the limit of vanishing collision energy. For the $X^2\Pi_{1/2}$ component, excitation of the rotational levels of $J=1/2, 3/2, \ldots 13/2$ are found. These excited fragments of the ion will shift the threshold energy
for ion-pair formation. The ion-pair limit has a threshold energy of 17 meV for the lowest $\Omega=3/2$ state and $J=3/2$ level. Note that even for rotationally relaxed ions, for the $\Omega=1/2$ component the ion-pair limit will be energetically open.

\begin{figure}
\rotatebox{0.0}{\includegraphics[width=1.0\columnwidth]{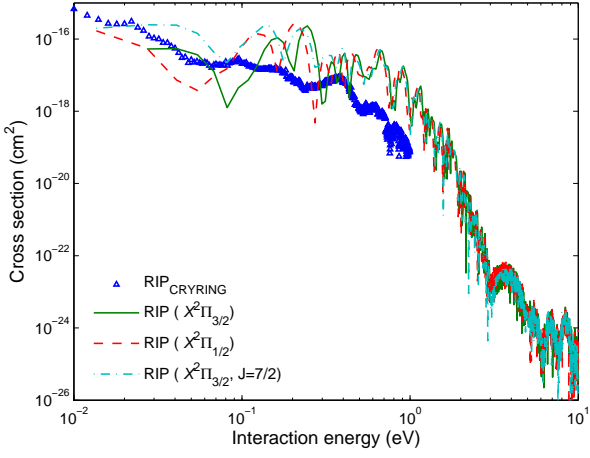}}
\caption{\label{fig:Vion1_Vion2} (Color online) Cross section for ion-pair formation in  electron recombination with a vibrationally and rotationally relaxed HF$^+$ ions in the $X^2\Pi_{3/2}$ [solid (green online)] or $X^2\Pi_{1/2}$ states [dashed (red online)]. The cross section for electron recombination with HF$^+$ in $X^2\Pi_{3/2}$ with $J=7/2$ is shown with the dashed-dotted (cyan online) curve. The triangles show the measured cross section from CRYRING~\cite{djuric01}.}
\end{figure}

In order to simulate the effect of ion-pair formation in electron recombination with the 
$X^2\Pi_{1/2}$ component of HF$^+$, a calculation is carried out where the ion-potential
is shifted upwards 30 meV relative to the neutral states. In figure~\ref{fig:Vion1_Vion2}
the calculated cross sections are compared with the measured cross section from CRYRING~\cite{djuric01}.
Since the ion potential is shifted upwards, less interaction energy is needed to reach the resonant state potentials and structures will therefore be shifted toward lower energies as can bee seen in the figure. 

Rotational excitation of the ion will also give rise to such effects. In order to calculate the ion-pair cross section of a rotationally excited ion such as HF$^+$ in $X^2\Pi_{3/2}$ state with $J=7/2$, not only the zero-point energy is shifted but also an effective term $J(J+1)/(2{\mu}R^2)$ is added to the diagonal elements of the potential matrix used in the wave packet propagation. The resulting cross section is shown Figure~\ref{fig:Vion1_Vion2}. 

In the measurements there is a distribution of the $\Omega$ components as well as the rotational excitation of the ion. This will cause 
the structures in the measured cross section to be washed out.
 This might help explain why less pronounced structures are found in the measured cross section compared to the theory.

\subsection {RIP of DF$^+$}
\begin{figure}
\rotatebox{0.0}{\includegraphics[width=1.0\columnwidth]{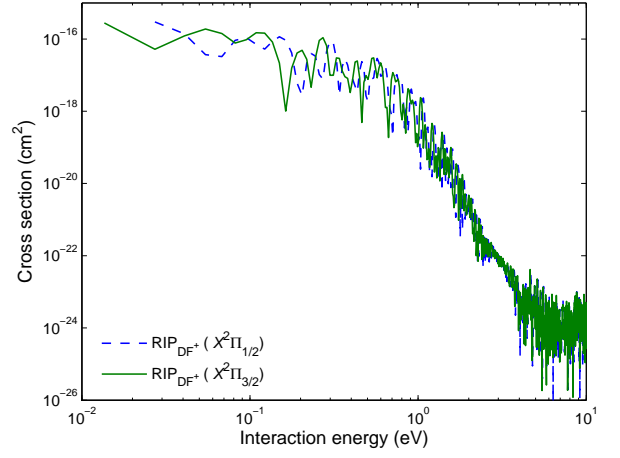}}
\caption{\label{fig:DF1_DF2} (Color online) Cross section for ion-pair formation in electron recombination with a vibrationally and rotationally relaxed DF$^+$ ions in the states $X^2\Pi_{3/2}$ or $X^2\Pi_{1/2}$.}
\end{figure}

By changing the reduced mass of the system when the vibrational wave function is calculated as well in the nuclear dynamics calculation, resonant ion-pair formation in electron recombination with DF$^+$ is studied. Again, model I is applied with 8 coupled neutral states.
The cross section for both $\Omega$ components of the ion, $X^2\Pi_{3/2}$ and $X^2\Pi_{1/2}$, are calculated and displayed in Figure~\ref{fig:DF1_DF2}.
The amplitude of the cross section of ion-pair formation in electron recombination with DF$^+$ is similar to that of HF$^+$, but the oscillating structures are in this case somewhat smaller. The cross section for ion-pair formation with DF$^+$ has not been measured.

\subsection{Total cross section in $^1\Sigma^+$ symmetry}
\begin{figure}
\rotatebox{0.0}{\includegraphics[width=1.0\columnwidth]{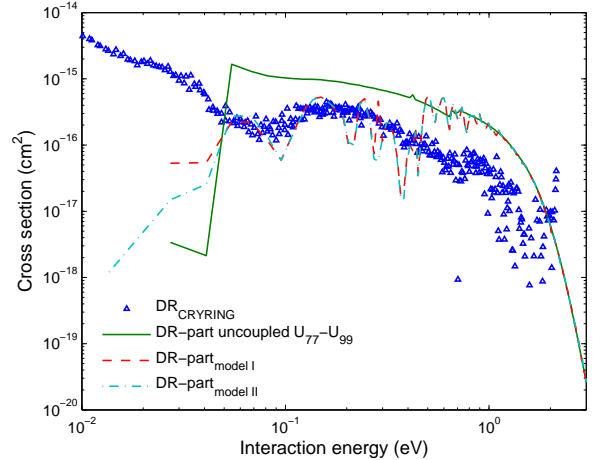}}
\caption{\label{fig:DR_part} (Color online) Contribution to the DR cross section from the three lowest resonant states of $^1\Sigma^+$ symmetry, calculated using uncoupled states, and the coupled models I and II. Also the measured cross section of DR~\cite{djuric01} is displayed.}
\end{figure}

The cross section for propagating wave packets on coupled states of $^1\Sigma^+$ symmetry is calculated using the two models I and II.
In order to study the effects of the couplings on the DR cross section, in
Figure~\ref{fig:DR_part}, the cross section for the coupled systems 
is compared with the cross section calculated by propagating wave packets on the three lowest uncoupled resonant states of $^1\Sigma^+$ symmetry. We also compare with the measured cross section for DR~\cite{djuric01}.

When the electronic couplings are included in the wave packet propagation, the total cross section for dissociation along the $^1\Sigma^+$ states is reduced in magnitude.
For low interaction energies, flux is lost into Rydberg states that are energetically closed for dissociation, making the magnitude lower in the energy range $0.044-0.5$ eV.
 
The cross section below the sharp threshold around 0.044 eV is somewhat increased when the
couplings are included. However, it is still several orders of magnitude smaller than the measured cross section. This might be explained by the existence of the excited $\Omega=1/2$ and rotationally excited ions in the experiments. Also, the indirect process, induced by the neglected non-adiabatic couplings between ionization continuum and Rydberg states might be important here.

Above $0.4$ eV, there is no longer a large difference in magnitude of the cross section calculated for the uncoupled and coupled systems. The couplings induce more structure in the total cross section. At even higher energies more resonant states not included in the present model become important. In our previous study~\cite{roos08}, 9 resonant states of $^1\Sigma^+$ symmetry were included. 

Also resonant states of other symmetries not considered in the present study, play an important role in DR of HF$^+$.
For the $^3\Sigma^+$ symmetry, we noticed~\cite{roos08} that the potentials as well as the partial cross sections were very similar to those of $^1\Sigma^+$ symmetry. The only large difference was that the lowest state of $^3\Sigma^+$ symmetry is associated with the ground state limit asymptotically. We can therefore expect a similar behavior if we introduce couplings between states of $^3\Sigma^+$ symmetry. The lowest states of $^1\Pi$ and $^3\Pi$ play a less important role in DR, since the partial cross sections were already lower than the experimental DR cross section in the whole energy range. 

\section{Conclusions}
We have calculated the electronic states of $^1\Sigma^+$ symmetry relevant for ion-pair formation in electron recombination of HF$^+$ using a combination of scattering calculations and MRCI structure calculations. We then propose a diabatization procedure, where not only the quasidiabatic potentials are obtained, but also the electronic couplings between the neutral states. The diabatization procedure is based on the construction of an orthogonal transformation matrix written as a product of successive two-by-two transformations.

The nuclear dynamics are described using wave packets that are propagated along coupled system including up to
nine states. Autoionization is included using the local approximation. With our models we  
obtain very good agreement in magnitude with the experimental cross section measured in CRYRING~\cite{djuric01}. Inclusion of the electronic coupling increases the cross section by about two orders of magnitude. Our calculated cross section shows structures similar to those in the measured cross section, but the structures are more pronounced and somewhat shifted relative to the experimental ones. 

The present study do not include spin-orbit coupling effects. To address these effects, the
ion-pair cross section for electron recombination with
the $X^2\Pi_{1/2}$ component of the ion, which lies about 30 meV higher in energy,
 is calculated. A cross section without a sharp threshold and shifted structures is obtained. Also recombination with
 rotationally excited ions ($J=7/2$ is calculated) which caused shifted structures in the cross section and produced a larger value of the cross
 section at low collision energies.
 
The ion-pair formation in electron recombination with DF$^+$ has also been calculated using the same potential energy curves, widths and couplings, but with a different reduced mass in the dynamical calculation.

By adding up the partial cross sections of the states of $^1\Sigma^+$ symmetry, included in the coupled system we examine the role of the couplings to the total cross section for DR of HF$^+$. We see clear indications of a reduced cross section for DR when electronic couplings are included. Better agreements with measured cross sections is obtained. \\

\begin{acknowledgments}
We thank A. Wolf for useful discussions and access to unpublished experimental data from his group.
\AA. L. acknowledges support from The Swedish Research Council and A. E. O. acknowledges support from the National Science 
Foundation, Grant No. PHY-05-55401.
\end{acknowledgments}

\end{document}